# American coot collective on-water dynamics

**Hugh Trenchard**

Correspondence address: Hugh Trenchard, 406 1172 Yates Street, Victoria, British Columbia htrenchard@shaw.ca; 1-250-472-0718

*Abstract: American coot (Fulica americana) flocks exhibit water surface (two-dimensional) collective dynamics that oscillate between two primary phases: a disordered phase of low density and non-uniform coot body orientations; a synchronized phase characterized by high density, uniform body orientations and speed. For this small-scale study, data was obtained for flocks of 10 to ~250 members for these phases. Approximate durations of phase occurrences were recorded for different flock sizes and for both relatively calm and severe weather conditions. Results indicate that for timed durations of up to ~ 2 hours, small flocks (10 coots) exhibit a comparatively high disordered/synchronized phase ratio (PR) in relatively warm and well-sheltered conditions (substantially >1); large flocks (~100 or more) generally exhibit a PR near 1, while large flocks spend comparatively more time in a disordered phase in relatively calm conditions (PR somewhat >1), and spend more time in a synchronized phase during severe conditions (PR <1). Data suggests a correlation between flock size and PR; and weather conditions and PR, whereby coupling principles driving on-water collective behavior include energy-savings and thermoregulation. Secondary phases occur, including expanding circle and sequential short distance flight, near-single-file lines, convex and concave arcs, among others.*

# INTRODUCTION

Pattern formation in biological collectives is well-studied, as are the dynamical principles that drive the pattern behavior (Vicsek & Zafiris, 2012 for overview). Fundamental to collective pattern formation is that patterns self-organize from the local interactions between individuals, causing different individual behaviors within the group than would be observed in isolation (Vicsek & Zafiris, 2012). Aggregate patterns that emerge from these collective interactions are unexpected and not possible at the level of the individual organism (Parrish & Edelstein-Keshet, 1999).

The study of collective motion in biological systems, described generally by Vicsek and Zafiris (2012) as flocking behavior, obtains analogies from the statistical physics of phase transitions of equilibrium systems, such as ferromagnets and liquid crystals (Czirok & Vicsek, 2000). Applying these analogies, flocking phases can be classified by their symmetries, including ferromagnetic (motion uniformity) and nematic (preferential alignment) symmetries (Toner, Tu & Ramaswamy, 2005). However, constituent organisms of biological systems are self-propelled and therefore exhibit non-equilibrium dynamics (Toner, Tu & Ramaswamy, 2005), and fall within the domain of complex adaptive, non-linear systems.

American coot collectives (*Fulica americana*) exhibit a variety of complex behaviors, both two-dimensional (on-water), and mixed two-dimensional and three-dimensional (flight), that appear driven primarily by the energy-savings and thermoregulatory benefits of optimal positions, and collision avoidance. I identify two main phases of coot flock on-water behavior, a synchronized phase and a disordered phase, and identify the main parameters of these phases, provide evidence for them, and identify a number of secondary phases of coot behavior.

Literature on American coot behavior includes extensive work on breeding, habitat and migratory behavior (McNair & Cramer-Burke, 2006), and most recently there have been contributions relating to brood parasite identification (Shizuka & Lyon, 2010) and fledgling success (Arnold, 2011). There appears to be little research examining coot water surface (two-dimensional) aggregate formations, and two-dimensional flock dynamics in general.

In one such study, however, Lukeman, Yue-Xian, and Edelstein-Keshet (2010) found that for the surf scoters studied, peak density occurred among birds directly in front or behind, a behavior replicated in a number of other animal collectives. Lukeman et al. (2010) show that on-water surf scoters are well-spaced, separated on average by 1.45 body lengths; and the authors observe that scoters maintain a zone of repulsion outside of which scoters endeavor to remain, and that this dynamic takes precedence over attractive and alignment dynamics.

By contrast, coots travel at angles to their direction of motion, and appear to come well within what is, for surf scoters, a zone of repulsion, as shown in Fig. 1. Hence, for coots, the collision avoidance dynamic is subordinate to the attractive and alignment dynamics.

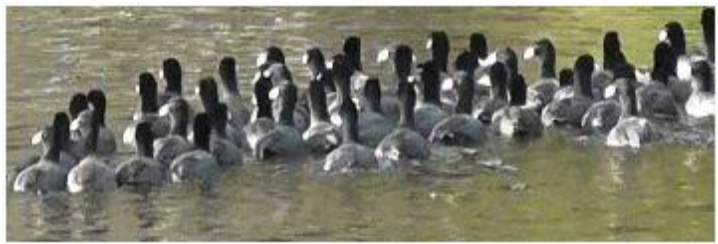

**Figure 1.** Coots in high density formation, clearly colliding with each other.

In this small-scale study, I observed flocks numbering between 10 and ~250 coots on Elk Lake, and Prospect Lake, approximately eight kilometres north of Victoria, on Vancouver Island, British Columbia. Observations occurred between late December, 2011, and mid-February, 2012. Observation periods were between one to three hours each, during which flocks swam on the water surface and spent time standing on docks and on the lakeshore. When surface swimming, coots usually covered regions of approximately 200m in diameter near the lakeshore. I took video footage using, on different occasions, a

Samsung H264 34X optical zoom hand-held camera, and a Panasonic HDC-HS80 42X digital zoom hand-held camera. Observations were also made without camera.

In addition to difficulties in identifying phase transitions, discussed in more detail in following sections, a major limitation of this study was the movement of flocks outside recording range, around lakeshore corners; their movement into lake areas where it was not practical to obtain footage or to the shore or rowing docks, resulting in discontinuous swimming time data collection. Also, on some days of attempted observation, I was unable to locate a flock. While I used flotation devices to locate and observe flocks later in the study period, data collection was at a very low angle, and generally lacked continuity due to required directional changes to the flotation device and inconsistent recording angles. Further issues included limitations in video battery life that truncated observation periods or resulted in discontinuous data.

## PRIMARY PHASES

Two primary behavioral phases were observed: a disordered phase of activity, as shown in Fig. 2a, and a synchronized phase, shown in Fig. 2b. The disordered phase was characterized by low-density, non-uniform body orientations and spontaneous darting and diving that appeared to occur randomly. The synchronized phase was characterized by high-density, uniform body orientation, well-defined geometrical patterns, and synchronized surface swimming speed.

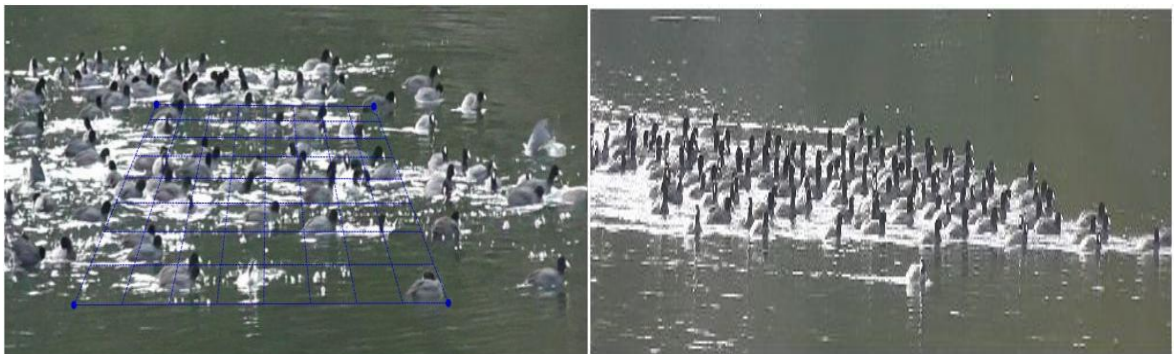

**Figure 2 (a).** Disordered phase, characterized by low-density, non-uniform shifting body orientations, diving, and darting. 8 X 8 grid indicates density of 0.51 c/g. **Figure 2(b).** Synchronized phase, characterized by high-density, uniform body orientation, speed, and travelling direction.

**Transitions Between Phases**

Two main factors appear to induce the transition from disordered to synchronized states: the first is acceleration among individual coots within the flock that brings accelerating coots in closer proximity to those moving more slowly; this induces slower moving coots to adjust their orientation into alignment with accelerating coots. The second is acceleration among individuals in lead group positions or on group peripheries that causes a widening separation between coots; this induces others to adjust their orientations and speeds in order to follow.

Time to transition for the whole group is proportional to density preceding the transition. This was most evident when synchronized, high density groups, shifted collective direction uniformly and almost instantaneously, compared to individuals within less dense groups which appeared to engage a greater number of orientation adjustments, possibly due to some indecision in choosing among neighbors to follow, requiring more time to synchronize their motion with neighbors.

The transition from order to disorder is indicated either by gradual decay of uniform orientation and high density by diving coots either at peripheries or within the cluster, or by sudden individual coot directional changes for foraging or other non-obvious purposes. After diving, coots frequently surfaced with vegetation in their beaks which other coots appeared sometimes to attempt to snatch from the foragers, causing sudden directional changes while a short chase ensued (~1 s), inducing a general breakdown of group cohesion. At other times, spontaneous breakdown in the synchronized state by flight occurred, as shown in Figures 6 – 8.

Notably, the onset of spontaneous order was more clearly demarcated and easily measured than was the onset of disorder. Except for instances of flight which suddenly destroyed on-water cohesion, onset of disorder was generally a continuous process. These transition periods, whereby states were briefly mixed and not well bounded, represent a problem in timing the duration of phases.

For two-dimensional coot flocks, phase transition parameters are changes in average velocity, density and orientation. Further, I suggest that, in combination with density, orientation, and velocity, coot flock on-water phases may be refined to include an average flock power output parameter, and ratios of power output of coots in position of highest drag (leading coots) and those in positions of lowest drag (following coots). This is a refinement of the velocity parameter because the energy savings benefits of travelling in zones of reduced hydrodynamic drag, discussed in more detail in following sections, do not correlate with speed; i.e. in high density states, coots reduce power output by following in zones of reduced water drag, meaning that as a high-density group they can swim faster by alternating leading positions, than when swimming in lower densities, or in isolation, at similar power outputs. This effect of increased group speed is shown, for example, to occur among cyclists in groups who share pace-setting (Olds, 1998).  Conversely, in high drag conditions (high wind and waves) coots may travel comparatively slowly at high power output even as a high density group.

In this study, for the purpose of timing disordered phase occurrences, I defined approximate phase boundaries according to greatest apparent proportion of coots visibly separated at proximities of at least one body length from each other and not in alignment or travelling at uniform speeds.  To time synchronized phase occurrences, I defined phase boundaries according to the greatest apparent proportion of coots aligned in orientation, and travelling at the same speed. Proximity between coots was a subordinate factor to orientation and speed when identifying the ordered phase, since uniform orientation and acceleration usually preceded increasing density in the synchronized phase; while decreasing density was more important in identifying the onset of disorder, since decreasing density often preceded or was simultaneous to breakdown in uniform orientation and speeds.

 Significant error is acknowledged in this approach to establishing phase boundaries and the timing of phase occurrences. The problem is exacerbated by optical angle distortion and coots' own blocking of external sightlines. This is solved accurately only by overhead observation, which was impractical for this study. While ladders were used in attempts to obtain views from as high up as possible to reduce sightline blocking and angle distortion, ladder utility was limited when flocks were moving locations continually.

Although it was possible to review video footage and estimate distances on the horizontal plane from left to right across photographic images using 37cm as an average coot body length (Alisauskas, R, 1987), it was not possible to accurately determine actual distances between coots on the vertical plane (front to back) toward the horizon due to optical angle distortion. Accurate measurements of flock dimensions are therefore not possible in this study.

In turn, accurate density measurement cannot be made in this study, since density is defined as $d = N/L^2$ (Czirok & Vicsek, 2000), where here $N$ is the number coots, and $L^2$ is the total two-dimensional area within which all birds are contained. Typical unit values for spatial area are in square metres (Beauchamp, 2012), or in body lengths (*BL*) squared (Cambui & Rosas, 2012).

However, as a method to compare flock densities within selected regions, I used Kinovea freeware to overlay an 8 X 8 grid over image regions such that the x-axis of one square in the grid corresponds to *BL* of one coot in the horizontal plane, oriented to the image perspective, as shown in Fig. 2a, and Fig. 3a. Where optical angles permit a reasonably accurate grid overlay, density for the region is indicated by the number of coots/64 unit grid (c/g). For Fig. 2a, a density 0.51 c/g is shown for the selected region. In Fig. 14 flock density is comparatively low, at 0.17 c/g. This density value is consistent with the lower extremes of flock density observed visually, although in Fig. 14 coots are not in a disordered state, as discussed in sections that follow. In Fig. 3a, flock density is 0.70 c/g for the region selected. In Figures 2b and 4, optical angles do not allow an accurate grid overlay, but if we apply $d=NL^2$ it is plain that density approaches 1.

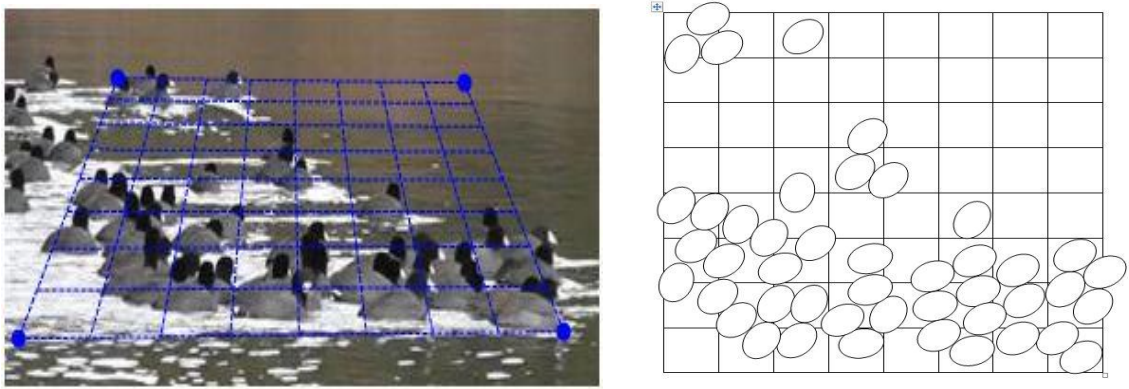

**Figure 3 (a).** Angle of alignment relative to orientation is difficult to determine given optical angle. Using grid overlay, positions and orientations are approximated and transformed to overhead view in **(b)** where there are 45 coots in the 64 square grid, where density = 0.70 c/g.

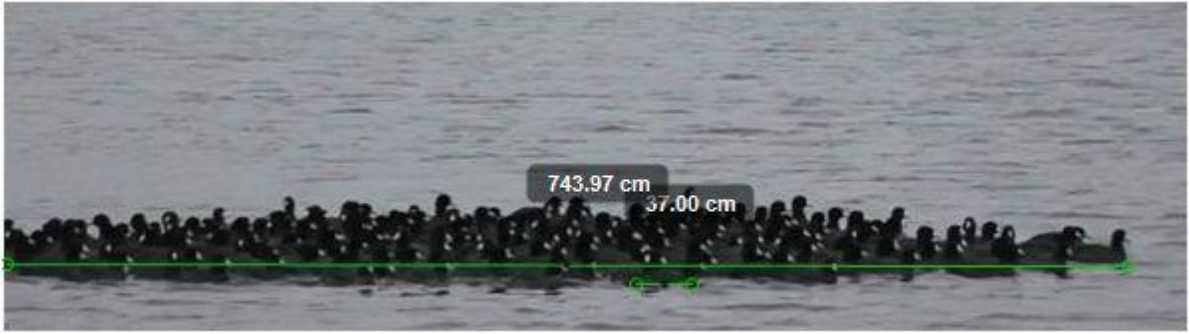

**Figure 4**. High density flock, synchronized in orientation and speed. Flock length is approximately 7.4 metres, based on 37cm average body length (Alisauskas, 1987). Estimated maximum y-axis width is 9 coots side by side; at 2/3 body length, the width is approximately 2.2 m. Applying $d=NL^2$, where body length is the unit of $L^2$, density approaches 1.

**Primary Phase Oscillations**

Transitions from disordered to ordered phases were indicated by a spontaneous increase in both density and alignment of orientation. Reverse transitions were indicated by decreasing density, frequent diving activity and spontaneous directional changes. Spontaneous breakdown of synchronized phases occurred when coots took flight in an expanding circle, as shown in Fig. 6. Transitions to a disordered phase appeared to be more gradual than transitions from disordered to ordered states.

Generally, the data here suggests that small groups tend to spend comparatively more time in disordered states, and the ratio of time spent by coots in disordered phases to time spent in synchronized phases ("phase ratio" or PR) falls with increasing group size to some unknown threshold size and PR. As shown in Table 1 and Fig. 5(a), for a flock of ten coots at Prospect Lake in well-sheltered conditions (although relatively windy) and ~8 °C, coots' PR was 6.74. A similar ratio was observed for a group of seven on another day. Compare this with the PR for much larger groups of ~250 coots and ~150 coots, on other days, as shown in Table 2 and Figures 5(b) and 5(c), in which PRs are near 1. Despite the noted problems in determining phase boundaries and the associated error in timing of phase durations, the substantial difference in PR for the small group of 10 coots compared to the narrower differences in PR

10for the data sets involving much larger groups, suggests a threshold group size sufficiently large for PR to narrow and approach 1.

The data also suggests some correlation between weather conditions and PR. As shown in Fig. 5(b), a group of ~250 coots, in relatively calm conditions exhibited a PR=1.42, on December 26, 2011. On January 22, 2012, a group containing fewer coots (~150) than observed on December 26, 2011, exhibited a PR=0.90, as shown in Fig. 5(c).

The respective PRs for these two days are not as widely different from each other than each is from the PR shown for February 11, 2012. The average disordered phase times of 1'59" (119 s) and 1'52" (112 s) respectively between the two days are similar, and so the significant difference in PRs between the two data sets of December 26, 2011, and January 22, 2012, is due primarily to the higher recorded durations spent in the synchronized phase. Despite the acknowledged error in determining phase boundaries and timing phase durations, the consistency in the average durations recorded for disordered phases between the two days supports the validity of the recordings of greater difference between the average synchronized phase timings between the two days, at 1'24" (84 s) and 2'04" (124 s), respectively. In turn, this supports the suggestion that there is a physical principle causing the difference.

In view of the comparatively high wind conditions and cool effective temperatures due to wind chill on January 22, 2012, this significant difference in PR may be due to the combined optimized effect of the high-density synchronized formation in increasing metabolic output by swimming, the insulating effect of huddling, and the reduction of power output in zones of reduced hydrodynamic drag.

As shown in Table 1, on January 22, 2012, mean wind speed was reported to be 37.97 km/h and gusted to 74.08 km/h, while temperatures ranged from 4.0 °C to 7° C (Almanac.com). On December 26, 2011, mean wind speed was 11.85 km/h and gusted to 46.30 km/h, while temperatures ranged from 5 °C to 8 °C (Almanac.com). Wind chill resulted in correspondingly lower effective temperatures on January 22, 2012, as shown in Table 1. On February 11, 2012, although sustained wind speeds were close to that of January 22, 2012, the effective temperature due to wind chill was similar to that of December 26, 2011. In addition, the area of observation at Prospect Lake was a bay well-sheltered from the wind by trees and higher embankments than the area of observation at Elk Lake, and so the effects of wind-chill were likely less than shown in Table 1.

**Table 1.** Weather Conditions for December 26, 2011, January 22, 2012, and February 11, 2012, at Victoria Harbour, approximately 8km from both Elk Lake and Prospect Lake. Weather data from www.Almanac.com, and wind chill calculated from www.easycalculation.com/weather/wind-chill.php.

|  | Dec. 26, 2011 | Jan. 22, 2012 | Feb. 11, 2012 |
|---|---|---|---|
| Mean wind speed (km/h) | 11.85 | 37.97 | 33.71 |
| Maximum wind gust (km/h) | 46.3 | 74.08 | 55.56 |
| Temperature range (°C) | 5.0 – 8.0 | 4.0 – 7.0 | 7.0 – 10.0 |
| Temperature mean (°C) | 6.3 | 5.2 | 8.1 |
| Wind chill at mean wind speed and mean temperature (°C) | 4 | 0 | 4 |
| Wind chill at max. gust and mean temperature (°C) | 1 | -2 | 3 |

In cold and windy conditions, birds must increase their metabolic rates if a constant body temperature is to be maintained (Goldstein, 1983). It is suggested that at some threshold combination of wind and effective temperature due to wind chill, coots must increase their metabolic rates to stay sufficiently warm for survival. An obvious means of increasing metabolic rates is for coots to increase their swimming time. Therefore, relative to time spent in the disordered phase when much of coots' activity is low-velocity movement, coots appear to spend more time swimming in a synchronized state to increase metabolism, while simultaneously optimizing energy expenditure and by swimming in zones of reduced hydrodynamic drag during on-water foraging endeavors (also inducing a corresponding increase in energy requirements), while also retaining body heat by the insulating effects of huddling.

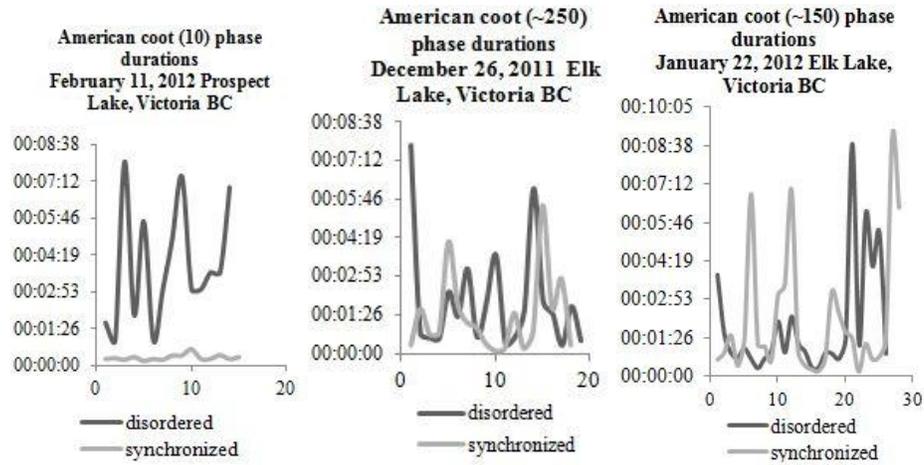

**Figure 5 (a-c).** Coot phase durations on three different observation days. Coot flocks alternated between disordered and synchronized phases. Each distinct phase occurrence was timed, and each plotted on separate curves to show relative phase durations. X- axis shows number of phase occurrences for each phase type, y-axis shows phase duration. See data summary in Table 2, below.

**Table 2**. Occurrences of Disordered and Synchronized Phases on Three Days, Average Time Coots Spent in Each phase, Total Time in Each Phase. Phase Ratio (PR) = average time (in seconds) of disordered phase / average time (in seconds) of synchronized phase. Note: difference in phase occurrence counts on January 22, 2012, is due to observations being made in two separate periods, each ending with a synchronized phase.

| | | Disordered Phase | Synchronized Phase | Total | Phase Ratio |
|---|---|---|---|---|---|
| Dec. 26, 2011 | Occurrences | 19 | 18 | | |
| | Average time of phase (min:sec) | 1:59 (119 s) | 1:24 (84 s) | | 1.42 |
| | Total time spent in phase (hr:min:sec) | 0:37:33 | 0:25:08 | 1:02:41 | |
| Jan. 22, 2012 | Occurrences | 26 | 28 | | |
| | Average time of phase (min:sec) | 1:52 (112 s) | 2:04 (124 s) | | 0.90 |
| | Total time spent in phase (hr:min:sec) | 0:48:30 | 0:57:58 | 1:46:28 | |
| Feb. 11, 2012 | Occurrences | 14 | 15 | | |
| | Average time of phase (mins:sec) | 3:56 (236 s) | 0:35 | | 6.74 |
| | Total time spent in phase (hr:min:sec) | 0:55:07 | 4:41 | 59:48 | |

## SECONDARY PHASES

Secondary phases are not easily categorized as either disordered on high-density synchronization, and exhibit some properties of both primary phases. Secondary phase durations were usually shorter than primary phases. The following describes some of these phases.

### Short Distance Flight Formations

Short-distance flight is defined here as air-travel in which coots' feet either touched the water surface or reached less than ~3 metres above water; the average duration of this phase was observed to be between 1 and ~15 seconds. Short-distance flight was observed in two primary situations: when groups reached a threshold density at which collisions occurred, causing cascading flight behavior, in either of two primary situations: more or less simultaneous flight, or sequential flight. Simultaneous flight was observed in expanding arcs or circles in high density formations, apparently to avoid collision, as shown in Fig. 6, and also in less dense approximate line formations, as shown in Fig. 8.

Sequential flight occurred generally from the exterior or posterior of the group, cascading in sequence inward toward the group centre or toward its anterior -- again, apparently in order to avoid collision, as shown in Fig.7. Sequential flight also occurred when lagging birds appeared to seek to integrate with a cluster ahead, as shown in Fig. 8; or, when birds sought to increase on-water speed by using their flight momentum upon landing and coasting.

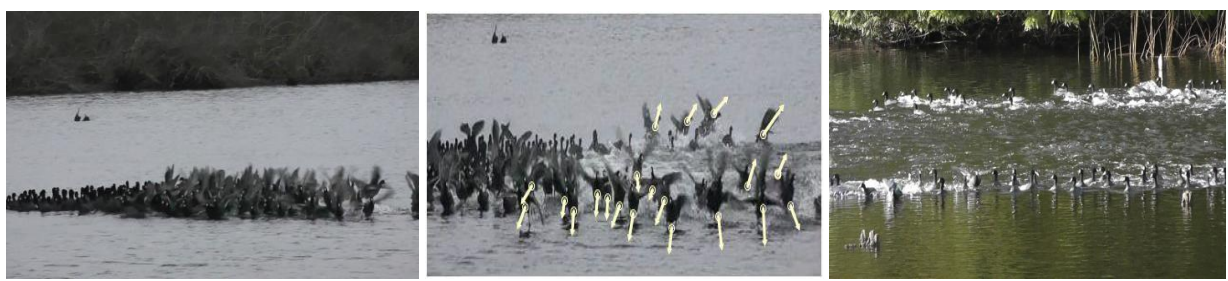

**Figure 6**. Three stages of flight dispersion. **(a)** Density has reached a critical threshold requiring density reduction. **(b)** Coots take flight away from each other in an approximate expanding circle to avoid collision. Arrows show approximate outward directions. **(c).** An example of a group landing after flight.

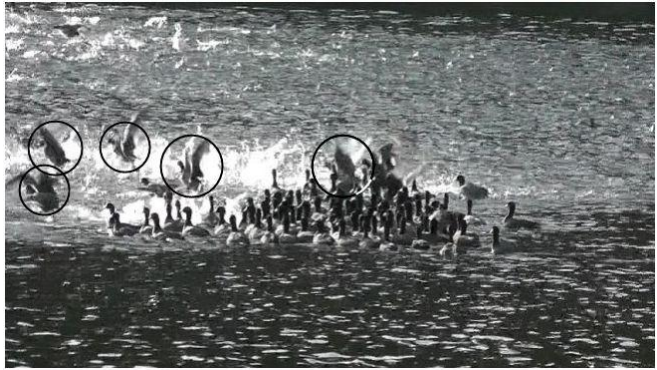

**Figure 7.** Sequential flight. Circles indicate birds first in flight, followed in sequence by birds at peripheries until all have taken flight.

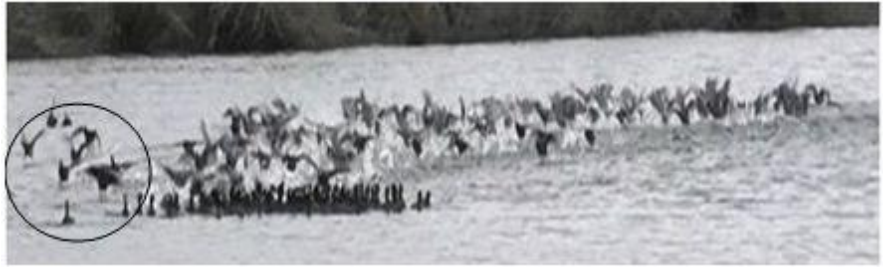

**Figure 8**. Sequential take-off among some coots, and simultaneous take-off among others. The apparent objective here is to "catch up" and integrate with the group ahead. Note birds within the drawn circle are turning to orient direction behind the main group on the water.

### Single-File Lines

Frequently observed were single-file formations, in which the entire group extends approximately one behind the other, with some overlapping, as shown in Fig. 9. Lukeman (2009) describes a similar dynamic among surf-scoters as a "follow-the-leader" line. During this process, coots regularly alternate anterior positions, whereby coots in positions two to three positions behind the lead coot will accelerate relative to the leading coot, and pass that coot to take the anterior position. This process is consistent with in-flight lead position alternations in the flight formations of other birds (Andersson and Wallander, 2004). It is also consistent with thermoregulation patterns observed in penguin huddles (Gilbert et al, 2006), and cooperative energy-savings patterns in bicycle pelotons (groups of cyclists) (Olds, 1998). Further, cyclists exhibit group-density oscillations, and exhibit single-file lines that appear to self-organize at critical power output thresholds (Trenchard, 2010), a dynamic Lukeman (2009) did not explore in relation to surf-scoter follow-the-leader lines. Applying these observations, it is hypothesized that coot rotational patterns and line formations emerge at critical metabolic thresholds as coots adjust following positions to maximize energy savings.

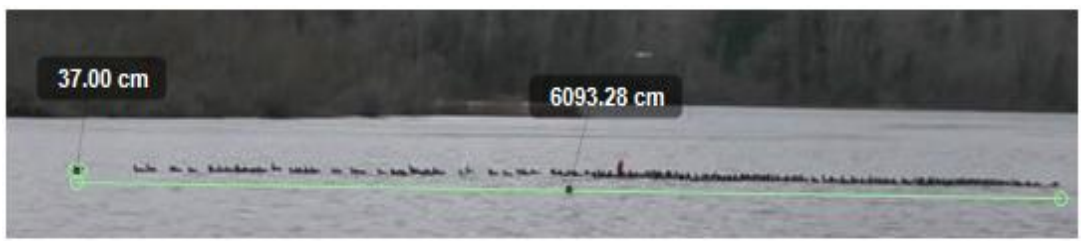

**Figure 9.** Line formation, approximately 60 meters long, synchronized in orientation and speed. Direction of travel toward left of image. Coots are aligned approximately single-file but line, with some overlapping from the near centre of the line backward (to image right).

### Arc Formations

Arcing formations occur with relative frequency, and appear to be partly a function of the angles to which coots align their bodies in close proximity, and their ovoid body dimensions. When arcing formations commence, the group expands outwards from initially higher-density formations, as shown in Figures 10(a) and 10(b), and appears as a convex arc as coots are oriented outward from the arc angle. To expand the arc, coots in central regions increase density to a threshold level, as in Fig. 10(b), which forces peripheral coots to advance further outward, while central coots adjust relative positions by drifting to the back and outside of the group. The pattern therefore emerges as a two-stage process: first, collective density increases to a threshold density, followed by positional adjustments outward to relieve this increasing density. Density freely increases in central regions due to a reduced drag coefficient relative to coots on peripheries. Note the difference between convex arcs resulting from this two-stage process, and concave arcs that result from the group diverging in two main directions with the coots oriented inward from the arc angle, as in Fig. 11, and arc formations that result from turning and changing their direction of travel, as in Fig. 12.

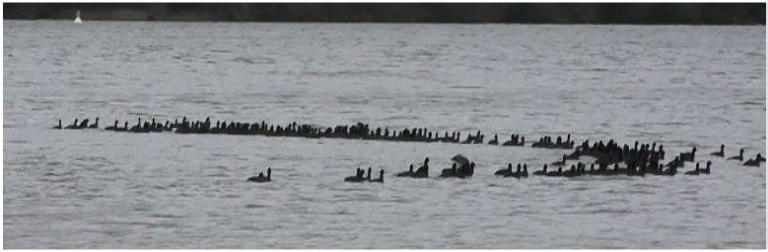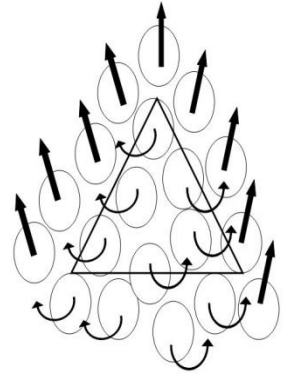

**Figure 10 (a).** Convex arc. Arcs appear to emerge as a result of high-density formations in which central regions increase density to a threshold, pushing peripheral coots outward, as depicted by straight arrows in **(b),** where the triangle indicates a zone of increasing density, as coots experience a lower drag coefficient relative to peripheral birds. Coots in high density central positions adjust relative positions back and outward to peripheral positions, pushing the arc outward, depicted by curved arrows.

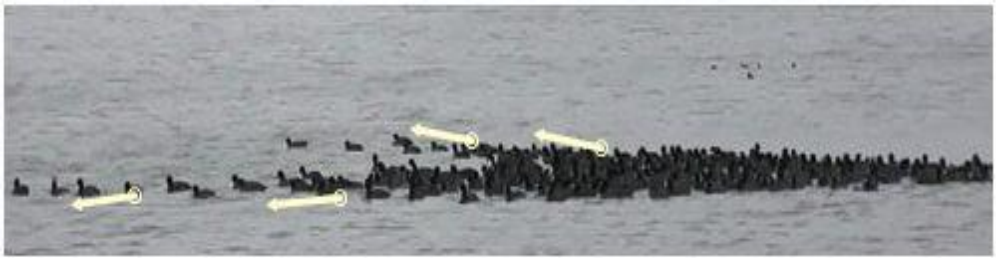

**Figure 11**. Concave arc in initial stages. Note two lines forming in different orientations, generating a concave arc as coots are oriented inward from the arc angle.

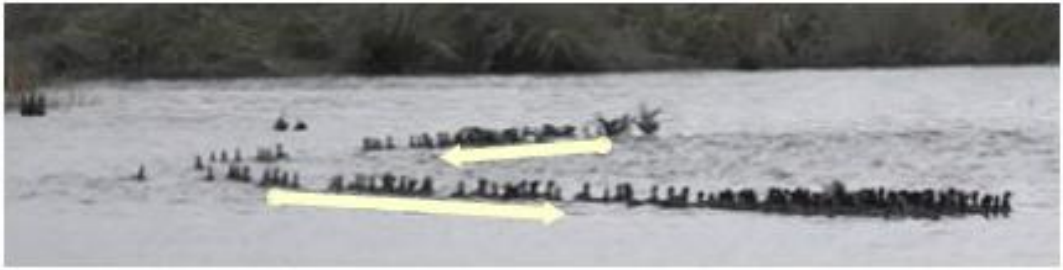

**Figure 12**. Turning formation. This is distinct from arc formations because coots are following along the same trajectory of motion, unlike arc formations in which trajectories are at small angles away from each other, as in Fig. 10(b). In this instance there was no attempt by coots to take the shortest distance across the arc to merge, which supports the hypothesis of optimized energy savings by following behind.

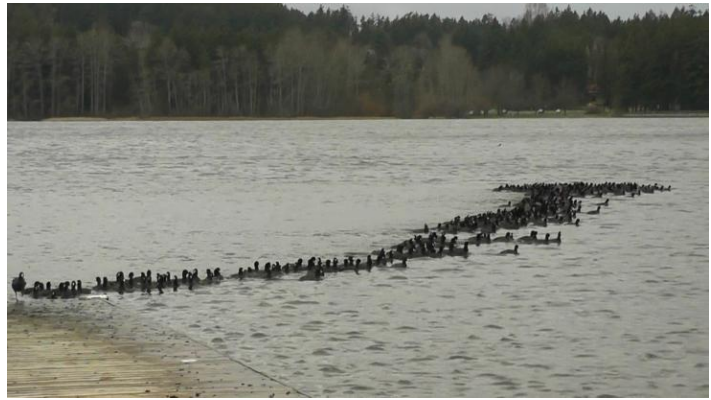

**Figure 13**. Laterally stretching group, and small angle arc. Coots approaching a dock indicate no fear of the photographer's presence within ~10m. Hence, predatory threats do not explain the high density formation here. Note the dock side on which a coot is standing is 5m wide (using Google Earth; full width not shown in this photo); coots were timed as they passed by the 5m dock side front at an approximate speed of 0.5 m/s.

## Low Density Synchronized States

Low-density uniform orientation states were observed, as shown in Fig. 14. In this formation, coots travelled at proximities outside optimal energy-savings zones, indicating that energy-savings is not a necessary precondition of synchronized locomotion. However, it is hypothesized that slower average speeds characterize the low density synchronized formation when compared with speeds in high-density synchronized formations, given similar weather and water conditions. However, further work is required to establish evidence for this hypothesis.

Conversely, high density disordered states were also observed. This is characterized by close-proximity diving and darting behavior. In conditions of sufficient density, darting motions in opposing directions led to the onset of milling, or vortex behavior, in which coots rotate en-masse in a circle. This behavior is common among animal aggregates (Lukeman, 2009).

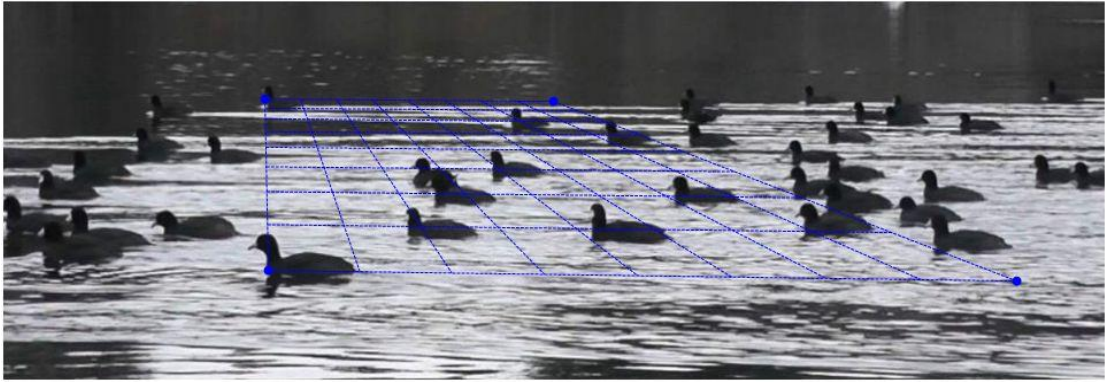

**Figure 14**. Low density synchronized phase. Coot speeds were not measured in this instance, but it is hypothesized that average group speed is lower in this type of low density state (in similar weather conditions) than when group density is higher, which allows coots to save energy in optimal positions of reduced drag. There are eleven coots in the 64 unit grid, for a density of 0.17 c/g.

## SELF-ORGANIZING PRINCIPLES

Drag reduction in nature by "drafting", or following behind others in optimal zones of reduced drag, is well established. Examples include bird flocks in flight (Weimerskirch, Julien, Clerquin, Alexandre & Jiraskova, 2001), fish schools (Weihs, 1973), dolphins (Weihs, 2004) and queues of spiny lobsters in water (Bill & Herrnkind, 1976).

Drag reduction among two-dimensional flocks is less well-studied. However, in one such study, Fish (1995) examined the energy savings of duckling water-surface formations. He observed that water vorticity generated by anterior (leading) ducklings provide momentum to the water in the same direction as the trailing ducklings. This is experienced by trailing ducklings as both reduced drag and energy expenditure; drag is further reduced by low spacing ratios. Fish (1995) found that optimal energy savings occurs in a single-file formation. He indicated additional reductions occur when a duck trails between two

19leading ducks swimming abreast, as in a diamond formation. Fish (1995) observed both single-file and diamond pattern formations for clutches of four ducks, and found that the least energetic effort is expended by the most posterior duckling in formation. He found that compared to a solitary duckling, ducklings trailing the mother in single-file had a 7.8-43.5% decrease in energy costs with increased savings for larger groups, including a maximum energy savings of 63% by four 3-day-old ducklings swimming in a decoy's wake compared to a solitary duckling.

Compared to ducks, coots differ in both body dimension and mass, and also possess three-toed lobed feet (Alisauskas, 1987), as opposed to the webbed feet of ducks. However, it is reasonable that substantially similar drag-reduction principles apply to coot surface swimming, and similar percentages of energy savings are obtained. As earlier discussed for the case of position alternations among cyclists in single-file lines and flying birds, in view of these energy savings benefits (power output reductions), it is reasonable that coots in non-optimal positions fatigue more quickly and decelerate relative to those in optimal power output positions. Fatiguing coots, in turn, are passed by "fresher" coots, thus inducing front position rotations. Some observations were made of positional rotations in broad planar formations (i.e. not single-file) in addition to rotations in single-file lines, as noted earlier, but further study is required to establish this principle as a primary cause of rotation dynamics.

Fish (1995) cautioned that there are other explanations for animal formation movement other than energy savings, including locating food resources, mating efficiency, pooling information, greater tolerance to toxic substances, and protection against predators (Parrish, J., Edelstein-Keshet, L., 1999).

In the case of American coots, locating food resources and energy savings are not mutually exclusive since there are high energy costs when swimming and feeding. Information pooling could well be an element of coot formations, but their high density formations reduce the capacity of coots in central regions to perceive information that could stimulate a physical response. The pooling of responses in the high density state is thus restricted to a proportionately small number of coots in the formation, likely those at the group's periphery. It is suggested here, therefore, that pooled information is a minor component of the synchronized phase state.

In terms of mating efficiency, the synchronized phase is inconsistent with mating behavior, since coots' mating process involves chasing behavior, more consistent with the disordered phase; and, actual copulation occurs generally near shorelines (Gullion, 1954), precluding synchronized behavior. Further, as Gullion (1954) observed, coots are monogamous, which suggests that high density water-swimming does not provide increased reproductive opportunities.

Ryan and Dinsmore (1979) suggested that coots, by swimming about their territory, may indicate that the territory is occupied and will be defended. However, similar to energy savings benefits in locating food resources, energy-savings may be exploited during territorial swimming, and so energy savings and territorial savings are not mutually exclusive.

I obtained anecdotal evidence that coots will spontaneously assume high density formations in the presence of eagles, which are known to attack coots (Sobkowiak, Titman, Sobkowiak, 1989). Similarly, I observed increases in collective density and speed in the presence of approaching boats, but this occurred only intermittently and accounts for a small fraction of my total observation time. However, over the entire scope of observation I observed no natural predators, and coots appeared generally unconcerned about my nearby presence, as shown in Fig. 13.

As further evidence that energy savings is the primary explanation for coot synchronized patterns, consider coot behavior in windy conditions. Hydrodynamic drag is increased due to waves generated by wind (Francis, 1950); and, obviously, aerodynamic drag is also increased. This supports the hypothesis that synchronized formations are driven partly by the energy savings benefits of drag reduction.

Further, as Goldstein (1983) has shown, in conditions of high winds, birds generally experience greater energy requirements when surface swimming. The increased time coots spent in the synchronized state (lower PR), during conditions of high wind and low effective temperature as shown in Fig. 5, suggests that synchronized flock behavior is also driven partly by energy savings due to thermoregulation when coots are subject to increased cooling, particularly in severe weather conditions. This synchronization is akin to the huddling behavior of penguins (Gilbert, Robertson, Le Maho, Naito, Anceli, 2006). Coots may increase body temperature in central regions of the group while simultaneously reducing energy expenditure, and rotate positions with cooler coots on the peripheries, allowing cooler coots to move to warmer interior regions of the flock.

In summary, to the extent that thermoregulation is a form of reduced energy consumption, obtained when body heat is conserved by huddling without increased individual metabolic requirement, I suggest that both close-proximity for drag reduction and collective thermoregulation are two components of a broader energy savings phenomenon.

## CONCLUSION AND DIRECTIONS FOR FURTHER RESEARCH

Collective American coot behavior represents a source of rich and extensive multi-dimensional research. The aim of this research has been to identify some of these behaviors and to examine the underlying principles that drive them. I have presented some data on oscillations between two primary phase states, a disordered state and a synchronized state, and I have argued that energy-savings and thermoregulatory benefits are the main principles of interaction from which high-density synchronized aggregate coot behaviors emerge.

Future research should include overhead footage by which to track positional movements. Data is required from different environmental conditions and among coot flocks of different size to establish correlations between physical conditions and the rich diversity of coot collective behavior. Computational models will assist this research.

A fluid dynamical model may be developed more fully. One component of a fluid dynamical model, viscosity, in the case of coots is a function of energy costly positional adjustments, which costs decrease initially as coots move into low-drag zones behind others. When group density increases to a certain threshold, viscosity begins to increase as coots collide with each other and increase energy expenditure to adjust their positions. When density and viscosity become too high, coots re-adjust positions spontaneously by taking flight, instantly reducing density and viscosity.

The critical density that precedes this kind of spontaneous flight may correspond to an optimal two-dimensional packing configuration. See Lopez and Beasley (2011) for current research on geometrical packing problems. The packing problem for American coots is defined by their ovoid body shape as they increase density at angles to their direction of motion. As observed, coots continue to increase density until a critical point at which stability is spontaneously broken by sequential flight or expanding circle flight.

Further, when individual coots alternate positions within a high density formation, the possibility of bio-convection arises. Bio-convection has been shown to occur among bacteria in water, such that bacteria swim to the water surface for oxygen and, as they accumulate at the surface, reach a threshold mass that is drawn down into the water by gravity (Kessler, Hoelzer, Pedley & Hill, 1994). It is suggested here that as coots increase density, they increase viscosity and friction by physical contact and positional adjustments that may result in speed reduction; or they may have to increase metabolic output to maintain speed while counteracting increased viscosity. At a critical threshold, it is predicted that a transition occurs to a convection roll pattern as coots alternate between high-viscosity internal regions and less viscous perimeter positions. Roll rotations are predicted to occur from internal/central regions out, or vice versa; or may occur as "plume rotations" (Kessler et al, 1994) as coots are squeezed in and out of perimeter positions. This is likely to occur in particularly severe weather conditions as over-heating coots seek external positions and cooling ones seek warmer internal positions. Studies should include metabolic and/or power output data, correlated with overhead video and motion tracking.